\documentclass{article}

\usepackage{graphicx} 
\usepackage{amsmath} 
\usepackage{subcaption}
\usepackage{amssymb}  
\usepackage{float}
\usepackage[most]{tcolorbox}

\usepackage{xparse}
\usepackage{natbib}
\usepackage{epstopdf}

\usepackage{xr}
\usepackage{hyperref}

\usepackage[accepted]{icml2019}

\def\exampletext{Example} 

\NewDocumentEnvironment{testexample}{ O{} }
{
\colorlet{colexam}{red!55!black} 
\newtcolorbox[use counter=testexample]{testexamplebox}{%
    empty,
    title={\exampletext: #1},
    attach boxed title to top left,
       minipage boxed title,
    boxed title style={empty,size=minimal,toprule=0pt,top=4pt,left=3mm,overlay={}},
    coltitle=colexam,fonttitle=\bfseries,
    before=\par\medskip\noindent,parbox=false,boxsep=0pt,left=3mm,right=0mm,top=2pt,breakable,pad at break=0mm,
       before upper=\csname @totalleftmargin\endcsname0pt, 
    overlay unbroken={\draw[colexam,line width=.5pt] ([xshift=-0pt]title.north west) -- ([xshift=-0pt]frame.south west); },
    overlay first={\draw[colexam,line width=.5pt] ([xshift=-0pt]title.north west) -- ([xshift=-0pt]frame.south west); },
    overlay middle={\draw[colexam,line width=.5pt] ([xshift=-0pt]frame.north west) -- ([xshift=-0pt]frame.south west); },
    overlay last={\draw[colexam,line width=.5pt] ([xshift=-0pt]frame.north west) -- ([xshift=-0pt]frame.south west); },%
    }
\begin{testexamplebox}}
{\end{testexamplebox}\endlist}

\newcommand{\DAG}{\textsc{Dag}}
\newcommand{\SRE}{\textsc{Sre}}
\newcommand{\gSEM}{\textsc{gSem}}

\newcommand{\mbs}[1]{\boldsymbol{#1}}
\newcommand{\basis}{\mbs{\phi}}
\newcommand{\basisr}{\mbs{\psi}}
\newcommand{\basissc}{\phi}
\newcommand{\T}{\top}

\newcommand{\resp}{y}
\newcommand{\expo}{z}
\newcommand{\cnf}{c}
\newcommand{\sps}{\mbs{s}}
\newcommand{\spssc}{s}

\newcommand{\bfunc}{f}
\newcommand{\ry}{w}
\newcommand{\rz}{v}
\newcommand{\effect}{\tau}
\newcommand{\effectest}{\widehat{\tau}}
\newcommand{\pvec}{\boldsymbol{\theta}}
\newcommand{\psc}{\theta}
\newcommand{\rvec}{\boldsymbol{\lambda}}
\newcommand{\rsc}{\lambda}
\newcommand{\del}{\boldsymbol{\delta}}
\newcommand{\dataset}{\mathcal{D}}

\DeclareMathOperator*{\argmin}{arg\,min}

\DeclareMathOperator{\E}{\mathbb{E}}

\DeclareMathOperator{\R}{\mathbb{R}}

\DeclareMathOperator{\col}{col}

\newcommand{\methodname}{\textsc{Rosce}}

\icmltitlerunning{Inferring Heterogeneous Causal Effects in Presence of Spatial Confounding}

\begin{document}

\twocolumn[
\icmltitle{Inferring Heterogeneous Causal Effects in Presence of Spatial Confounding}



\icmlsetsymbol{equal}{*}

\begin{icmlauthorlist}
\icmlauthor{Muhammad Osama}{uu}
\icmlauthor{Dave Zachariah}{uu}
\icmlauthor{Thomas B. Sch\"{o}n}{uu}
\end{icmlauthorlist}

\icmlaffiliation{uu}{Division of System and Control, Department of Information Technology, Uppsala University}

\icmlcorrespondingauthor{Muhammad Osama}{muhammad.osama@it.uu.se}
\icmlcorrespondingauthor{Dave Zachariah}{dave.zachariah@it.uu.se}

\icmlkeywords{Machine Learning, ICML}

\vskip 0.3in
]



\printAffiliationsAndNotice{\icmlEqualContribution} 

\begin{abstract}
We address the problem of inferring the causal effect of an exposure on an outcome across space, using observational data. The data is possibly subject to unmeasured confounding variables which, in a standard approach, must be adjusted for by estimating a nuisance function. Here we develop a method that eliminates the nuisance function, while mitigating the resulting errors-in-variables. The result is a robust and accurate inference method for spatially varying heterogeneous causal effects. The properties of the method are demonstrated on synthetic as well as real data from Germany and the US.

\end{abstract}

\section{Introduction}

In a wide range of scientific analyses, the quantity of interest is the causal effect of an exposure $\expo$ on an outcome $\resp$, which may vary across space. Consider, for instance, the causal effect of fertilizer usage $\expo$ on crop yield $\resp$. The effect, denoted $\effect$, quantifies how the average crop yield changes per unit of change of fertilizer quantity. Due to variations in the soil, $\effect$ will vary as a function of spatial location $\sps$. 

\begin{figure}
    \centering
    \begin{subfigure}{0.45\linewidth}
    \includegraphics[width=0.9\linewidth]{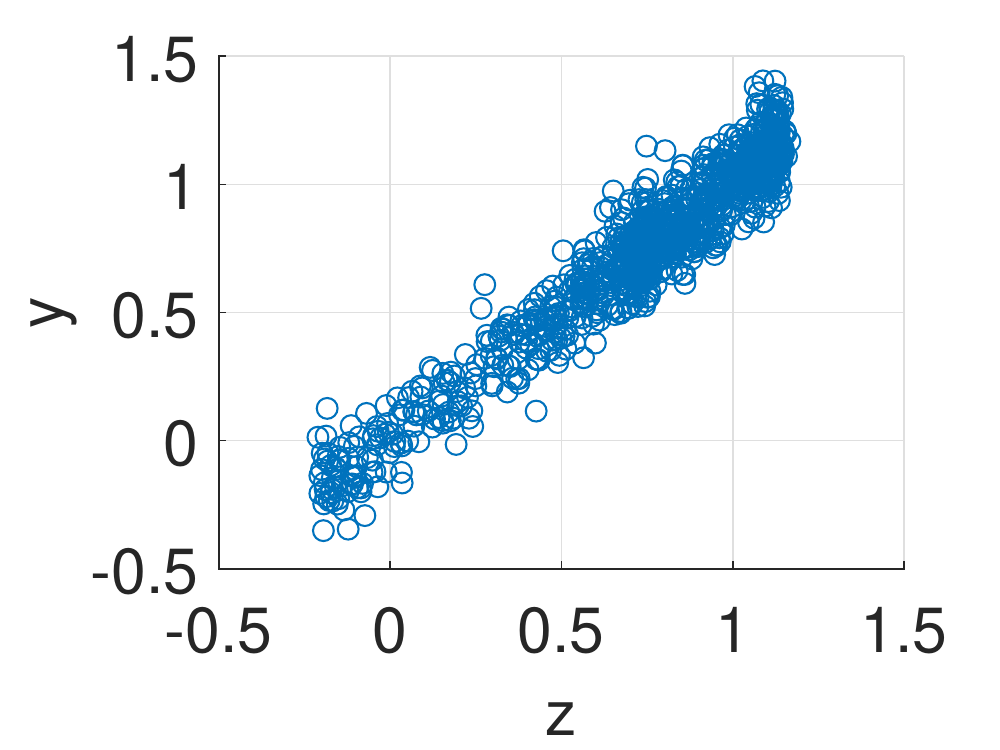}
    \caption{}
    \label{no_effect_scatter}
    \end{subfigure}
    \begin{subfigure}{0.45\linewidth}
    \includegraphics[width=0.9\linewidth]{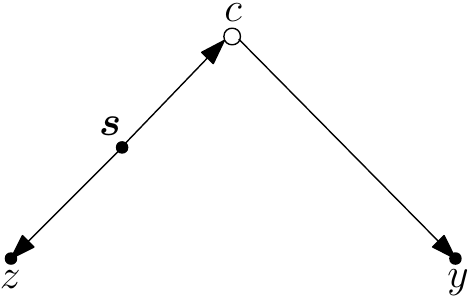}
    \caption{}
    \label{dag_no_effect}
    \end{subfigure}
    
    \caption{Illustration of spatial confounding. (a) Data exhibiting a clear association between exposure $\expo$ and outcome $\resp$. (b) Causal structure underlying the data-generating process, summarized as a \DAG{} where $\sps$ denotes spatial location and $\cnf$ is an unmeasured variable. Note that the causal effect of the exposure is zero.}
    \label{problem setup}
\end{figure}

The goal in this paper is to infer  the effect $\effect$ from a finite dataset $\dataset_n~=~\{(\resp_i,\expo_i,\sps_i)\}_{i=1}^{n}$. Unmeasured confounding variables, however, render this problem non-trivial. Consider a case illustrated in Figure~\ref{no_effect_scatter}, where there is a positive association between exposure and outcome yet the causal effect is zero, i.e., $\effect \equiv 0$. The causal structure underlying the data-generating process can be expressed using structural equations and summarized by a directed acyclic graph (\DAG) \cite{peters2017elements,pearl2009causality,hernan2010causal}. The structure is illustrated in Figure \ref{dag_no_effect}, where 
$\cnf$ denotes an unmeasured confounding variable which gives rise to a strong association between $\expo$ and $\resp$ despite there being no causal effect. This situation is referred to as \emph{spatial confounding}.

The bulk of the existing literature considers a fixed effect $\tau$ across space and, moreover, views $\cnf$ as a residual spatial variation using a spatial random effect (\SRE) model \cite{cressie1992statistics}. For example, \citet{reich2006effects} estimate the effect of socioeconomic factors on stomach cancer incidence ratio using a Gaussian prior model for $\cnf(\sps)$. Such approaches end up biasing the effect estimates and  various restricted spatial regression techniques attempt to mitigate these problems, cf. \cite{hodges2010adding,hughes2013dimension,hanks2015restricted,paciorek2010importance,page2017estimation}. The geoadditive structural equation model (\gSEM) in \citet{thaden2018structural} eliminates the bias problem by exploiting the causal structure underlying the data generating process. Specifically, they employ a linear Gaussian structural equation model and implicitly remove the spatial variation in $\expo$ and $\resp$ due to $\cnf$.
While it is capable of removing systematic errors, the \gSEM{} method is only applicable to discrete space and homogeneous or fixed causal effects. The effect may, however, be \emph{heterogeneous} with respect to spatial location. For example, when applying the method to study the effect of work experience on monthly income \cite{thaden2018structural}, it is quite possible that experience will have greater effect in regions with dense industries and high-skilled jobs, as compared to areas dominated by low-skilled jobs.

Existing methods for heterogeneous causal effect build upon statistical machine learning methods but do not address spatially varying effects specifically and are, moreover, mainly restricted to binary or categorical exposures $\expo$ \cite{imai2013estimating,wager2017estimation,kunzel2017meta,nie2017quasi}. The standard formulation of the inference problem requires estimating a nuisance function, which leads to significant biases when inferring $\effect$ even in the large-sample regime. The orthogonalized formulation introduced by \citet{chernozhukov2016double} circumvents this problem but results in multiplicative errors that are not addressed in the finite sample regime.

In this paper, we develop an inference method based on the orthogonalized formulation that
\begin{itemize}
    \item can infer heterogeneous causal effects with respect to spatial location,
    \item is operational in discretized as well as continuous space,
    \item provides robust and conservative estimates in the finite sample regime.
\end{itemize}
The outline of the paper is as follows: In Section \ref{prob. form}, we present the problem setup and define the heterogeneous effect. In Section \ref{method}, we introduce a model of the effect and develop a robust orthogonalized formulation. The resulting method is demonstrated on both simulated as well as real data in Section \ref{experimental results}. 

\textbf{Notation:} $\col\{a_1,a_2\}$ stacks both elements into a single column vector. $\otimes$ denotes the Kronecker product. $\E_n[a]~=~\frac{1}{n}\sum_{i=1}^{n}a_i$ denotes the sample mean of $a$.  

\section{Problem Formulation} \label{prob. form}

We consider continuous exposures $\expo$ and outcomes $\resp$. Spatial location or region is indexed by $\sps\in\mathcal{S}$, where $\mathcal{S}\subset \R^d$, if space is continuous, or $\mathcal{S}=\{1,\ldots,d\}$, when space is partitioned into $d$ discrete regions.

\begin{figure}[!t]
    \centering
    \begin{subfigure}{0.40\linewidth}
    \includegraphics[width=0.9\linewidth]{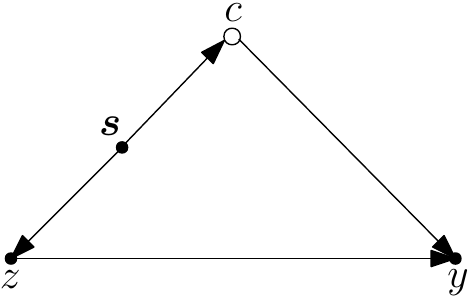}
    \caption{}
    \label{dag_1}
    \end{subfigure}
    \begin{subfigure}{0.40\linewidth}
    \includegraphics[width=0.9\linewidth]{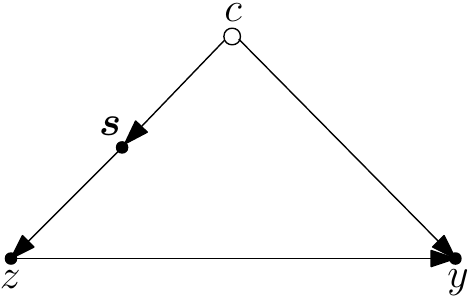}
    \caption{}
    \label{dag_2}
    \end{subfigure}
    \begin{subfigure}{0.40\linewidth}
    \includegraphics[width=0.9\linewidth]{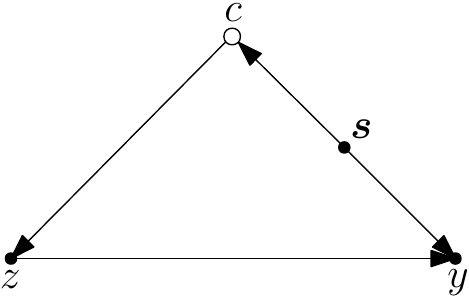}
    \caption{}
    \label{dag_3}
    \end{subfigure}
    \begin{subfigure}{0.40\linewidth}
    \includegraphics[width=0.9\linewidth]{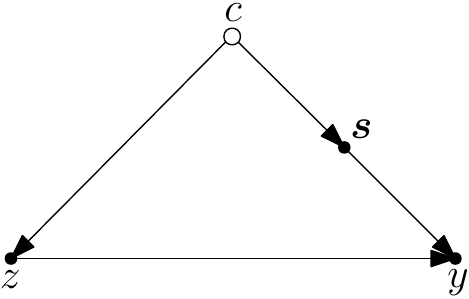}
    \caption{}
    \label{dag_3}
    \end{subfigure}
    \caption{Examples of different $\DAG$ configurations that result in spatial confounding, where $\cnf$ is an unmeasured confounding variable. The proposed method can address scenarios that include (a)-(d).}
    \label{dag confg}
\end{figure}

Let $\resp(\widetilde{\expo})$ be the counterfactual outcome had the exposure been assigned to $\widetilde{\expo}$, contrary to the observed value $\expo$. The (average) causal effect of the exposure at location $\sps$ is then defined as 
\begin{equation} \label{eq1}
\boxed{\tau \: \triangleq \: \frac{d}{d\widetilde{\expo}}\E\big[ \: \resp(\widetilde{\expo}) \: \big|\: \sps \: \big],}
\end{equation}
We consider scenarios in which the conditional independence
\begin{equation}
\expo \perp \resp(\widetilde{\expo}) \: | \: \sps   
\label{eq:conditionalindependence}
\end{equation}
holds. Figure~\ref{dag confg} shows examples of causal structures for which \eqref{eq:conditionalindependence} is valid. To illustrate the effect of an unmeasured confounding variable, we consider a simple example based on the structure in Fig.~\ref{dag_1}. 



\begin{figure*}[!h]
    \centering
    \begin{subfigure}{0.3\linewidth}
    \includegraphics[width=0.9\linewidth]{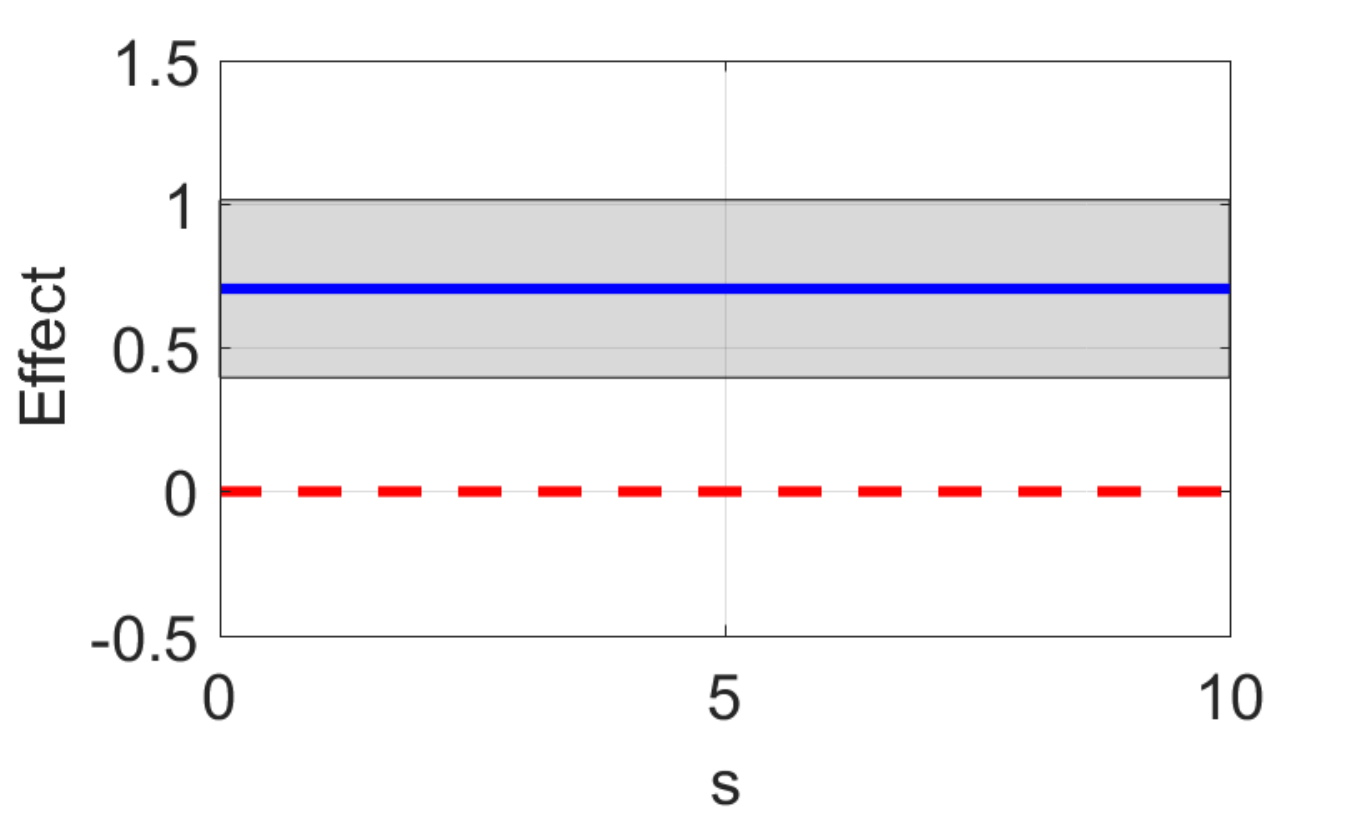}
    \caption{}
    \label{fixed_effect_rand_effect_model}
    \end{subfigure}
    \begin{subfigure}{0.3\linewidth}
    \includegraphics[width=0.9\linewidth]{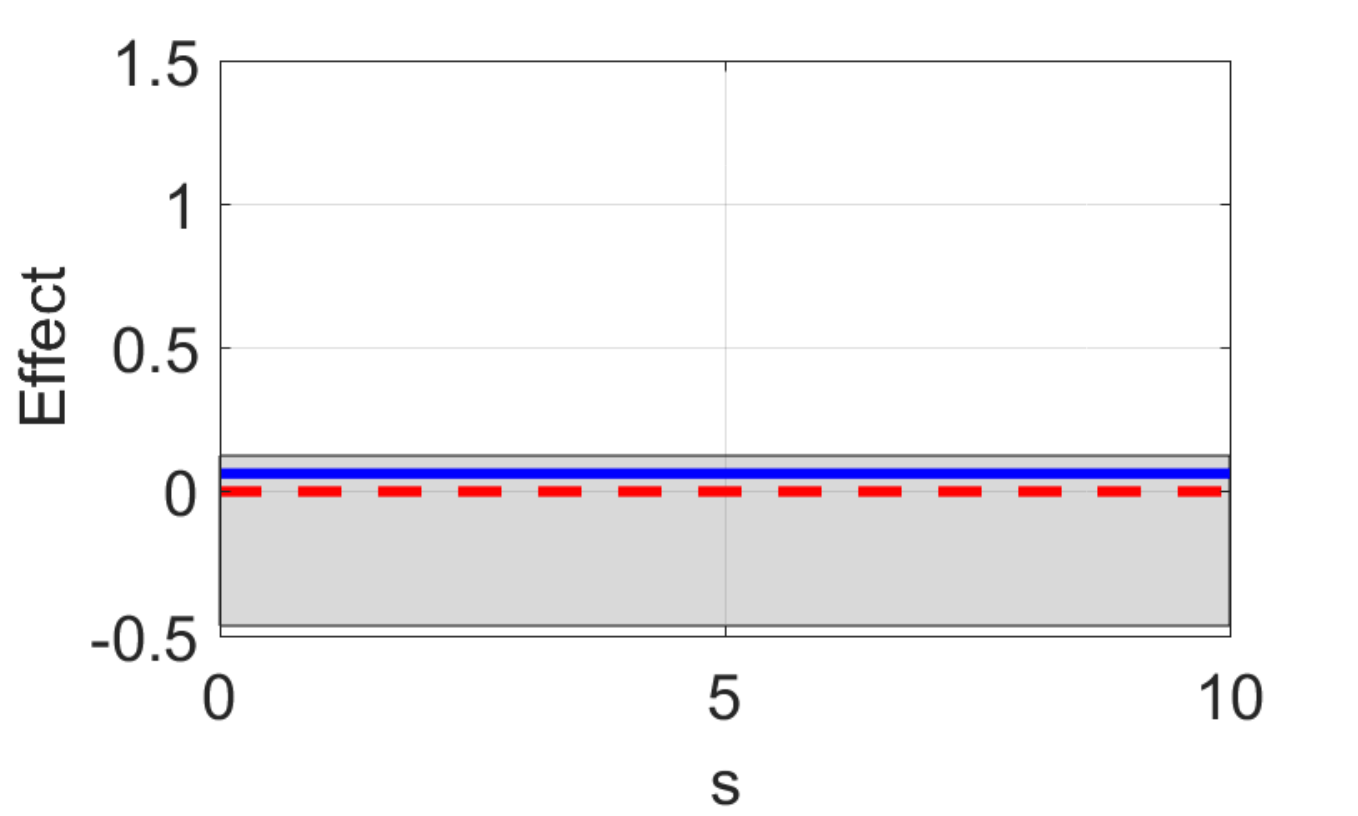}
    \caption{}
    \label{fixed_effect_our_model}
    \end{subfigure}
    \begin{subfigure}{0.3\linewidth}
    \includegraphics[width=0.9\linewidth]{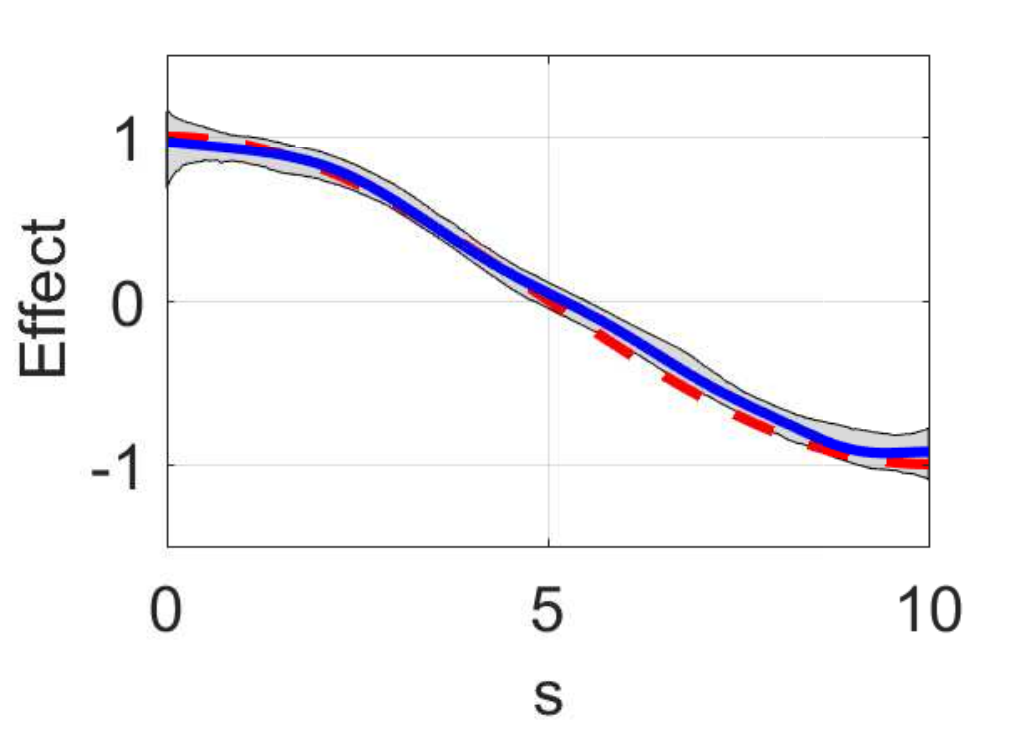}
    \caption{}
    \label{cont_var_effect_our_model}
    \end{subfigure}
   
    \caption{Illustration of example. The effect $\effect(\sps)$ of the exposure (dashed line) and its estimates $\widehat{\effect}(\sps)$ (solid line) with  95\% CI (shaded regions). Case 1: using (a) the \SRE{} model and (b) The proposed method. Case 2: proposed method in (c). Note that the $\SRE$ model is not applicable in this case.}
    \label{example}
\end{figure*}

\begin{testexample}
Let space $\sps$ be continuous one dimensional. We generate data $\mathcal{D}_n$ shown in Fig.~\ref{no_effect_scatter} based on the structure 
\begin{align*} \label{dgp_example}
    \resp&=\tau(\sps)\expo+c(\sps),\\
    \expo&=c(\sps)+\epsilon,
\end{align*}
which corresponds to Fig.~\ref{dag_1} and where $c(\sps)$ is an unmeasured confounding variable described as a spatial Gaussian Process
\begin{equation*}
    c(\sps)\sim \mathcal{GP}(0,k(\sps,\sps'))
\end{equation*}
with a Mat\'{e}rn plus white noise covariance function \cite{williams2006gaussian} and $\epsilon$ is white Gaussian noise. We consider two cases of effects:\\
\textbf{Case 1: Fixed effect $\tau(\sps)~\equiv~0$}. A standard approach is to parameterize the unknown effect by a constant and adjust or control for the unmeasured $c(\sps)$ by considering it as spatially structured noise \cite{cressie1992statistics}. The effect is then readily estimated using a generalized least squares method that weights the residuals based on the covariance structure of $c(\sps)$. Fig.~\ref{fixed_effect_rand_effect_model} shows the resulting estimate $\widehat{\tau}$ and its $95\%$ confidence interval (CI) along with $\effect(\sps)$. As can be seen, the estimate is biased and provides misleading inferences. By contrast, the method we propose in this work produces an estimate with $95\%$ CI that clearly contains the true effect, see Fig.~\ref{fixed_effect_our_model}.
\\  
\textbf{Case 2: Heterogeneous effect $\tau(\sps)$}. In some applications, the effect may vary significantly across space and even change sign. Unlike the standard approaches, the proposed method can also estimate spatially varying $\effect(\sps)$, as shown in Fig.~\ref{cont_var_effect_our_model}. The estimate closely follows the true effect, whose sign changes across space.
\end{testexample}

The inability to control for unmeasured $c(\sps)$ by exploiting its spatial correlation structure arises because the variable is taken to be independent of the exposure $\expo$ \cite{paciorek2010importance}. As we see in the example, however, $\expo$ is highly correlated with $\cnf(\sps)$.

We now formulate the problem in the general case, building upon \eqref{eq:conditionalindependence} and modularity \cite{pearl2009causality,richardson2013single,hernan2010causal} we obtain the identity
\begin{equation} \label{eq2}
   \E\big[\resp(\widetilde{\expo})|\sps\big] \:  \equiv \: \E\big[\resp|\expo=\widetilde{\expo},\sps\big].
\end{equation}
Locally at $\sps$, we assume that the expected outcome \eqref{eq2} is well approximated by an affine function with respect to the exposure $\widetilde{\expo}$. Then it follows from \eqref{eq1} that the effect is a function of space alone, that is, $\tau(\sps)$. By combining \eqref{eq1} and \eqref{eq2}, we get
\begin{equation*}
        \effect(\sps)=\frac{d}{dz}\E[\resp|\expo,\sps]\bigg\rvert_{\expo=\widetilde{\expo}}
\end{equation*}
After integrating both sides with respect to $\expo$, this yields
\begin{equation}
\E[\resp|\expo,\sps]=\effect(\sps)\expo+\beta(\sps)
\label{eq:conditionalresp}
\end{equation}
which implies the following form of the outcome
\begin{equation} \label{eq3}
    \resp=\effect(\sps)\expo+\beta(\sps)+\varepsilon,
\end{equation}
where $\beta(\sps)$ is a nuisance function and $\varepsilon$ is a zero-mean error term that is uncorrelated with any function of $\expo$ and $\sps$. We see that in the general case, $\beta(\sps)$ will be correlated with a spatially varying exposure $\expo$. Therefore omitting it, as in the example above, gives rise to bias due to spatial confounding. On the other hand, taking this term directly into account requires the estimation of a high-dimensional nuisance function $\beta(\sps)$ jointly with the effect $\effect(\sps)$. This leads to non-negligible biases in the finite sample case as pointed out by \citet{chernozhukov2016double} (Illustrated in Fig~\ref{2D_cont} below). These limitations motivate an alternative approach to infer $\effectest(\sps)$ from $\dataset_n$.

\section{Method} \label{method}
In this section, we begin by deriving an orthogonalized formulation of the problem that circumvents the need to directly estimate the nuisance function $\beta(\sps)$ in \eqref{eq3}. Then we develop a robust errors-in-variables method that yields heterogeneous effect estimates $\widehat{\effect}(\sps)$ as a function of the spatial location $\sps$. Finally, we specify a spatial predictive model that we employ in the subsequent experimental section.

\subsection{Orthogonalized formulation}
Consider residualized versions of the outcome and exposure, $\ry~=~\resp-\E[\resp|\sps]$ and $\rz~=~\expo-\E[\expo|\sps]$, respectively. These residuals are orthogonal to $\sps$ and in combination with \eqref{eq:conditionalresp} and \eqref{eq3}, we obtain an orthogonalized formulation
\begin{equation}
\begin{split}
    \ry&= y - \int \E[\resp|\expo,\sps] p(\expo|\sps)\: d\expo \\
    &=\effect(\sps)\rz+\varepsilon,
\end{split}
\label{eq5}
\end{equation}
which eliminates the nuisance function $\beta(\sps)$. 

To estimate the effect, which is a general function of $\sps$, we consider a flexible parametric model class
\begin{equation} \label{eq:parametricmodel}
    \effect_{\psc}(\sps) \in \big\{ f(\sps) : f= \basis^\T(\sps)\pvec \big\},
\end{equation}
where $\basis(\sps)$ is a spatial basis vector described below and $\pvec$ is a $d_{\psc}\times1$ parameter vector.  Thus the effect is directly identifiable from the residuals given in \eqref{eq5}.

\subsection{Robust errors-in-variables approach}
In practice, however, the conditional means $\E[\resp|\sps]$ and $\E[\expo|\sps]$ are unknown and they are replaced by empirical predictors. Thus the residuals can be written as
\begin{subequations} \label{residuals}
\begin{align}
\ry &= \underbrace{\big(\resp - \widehat{\E[\resp|\sps]} \big)}_{\widehat{\ry}} \; + \; \underbrace{\big( \widehat{\E[\resp|\sps]}- \E[\resp|\sps] \big)}_{\widetilde{\ry}} \\
\rz &= \underbrace{\big(\expo - \widehat{\E[\expo|\sps]} \big)}_{\widehat{\rz}} \; + \; \underbrace{\big(\widehat{\E[\expo|\sps]}- \E[\expo|\sps] \big)}_{\widetilde{\rz}} ,
\end{align}
\end{subequations}
where $\widetilde{\ry}$ and $\widetilde{\rz}$ denote errors. Let  
\begin{equation}
    \del(\sps) = \widetilde{\rz}\basis(\sps)
\label{eq:del}
\end{equation}
denote an unobservable random vector. Plugging \eqref{residuals} and the parametric model from \eqref{eq:parametricmodel} into \eqref{eq5}, we obtain the following relation
\begin{equation}
    \widehat{\ry}=\big(\widehat{\rz}\basis(\sps)+\del(\sps)\big)^\T\pvec \; + \; \widetilde{\varepsilon},
    \label{eq:residualregression}
\end{equation}
where $\widetilde{\varepsilon} = \varepsilon - \widetilde{\ry}$ is an unknown error. Hence using \eqref{eq:residualregression}, it is therefore possible to estimate the effect as
\begin{equation} \label{effect estimate}
    \widehat{\effect}(\sps)=\basis^\T(\sps)\widehat{\pvec},
\end{equation}
where the parameter vector $\widehat{\pvec}$ is learned from $\widetilde{\dataset}_n = \Big\{ \big( \widehat{\ry}_i, \widehat{\rz}_i, \basis(\sps_i) \big) \Big\}^n_{i=1}.$ This can be done using the least-squares (LS) method, which ignores the unknown deviation $\del(\sps)$ in \eqref{eq:residualregression}. It was shown by \citet{chernozhukov2016double} that LS nevertheless has good asymptotic properties in the case of fixed effects, that is when $\effect_{\psc}(\sps) \equiv \psc$. The deviation $\del(\sps)$, however, yields errors that are correlated with $\pvec$ so that the LS-based estimate exhibits notable finite-sample biases.

To tackle this problem, we propose using a robust learning method that takes into account the errors-in-variables arising from the unknown deviations. Specifically, we formulate a convex minimax problem
\begin{equation} 
    \widehat{\pvec} = \argmin_{\pvec}\left\{\max_{\del \in \Delta} \: \sqrt{\E_n\Big[ \big| \widehat{\ry} - \big(\widehat{\rz}\basis(\sps)+\del\big)^\T\pvec \big|^2 \Big]}  \:  \right\},
\label{eq:min-max-opt}
\end{equation}
where $\Delta$ denotes an uncertainty set for a deviation $\del$. In this way the method mitigates the worst case deviations in the set, which we now determine on conservative statistical grounds.

When using an empirical predictor of the exposure, it is reasonable to assume that the error term $\widetilde{\rz}$ in \eqref{residuals} tends to be smaller than $\widehat{\rz}$, in terms of their second-order moments. Considering the unknown deviation \eqref{eq:del}, this assumption on $\widetilde{\rz}$ motivates a conservative upper bound on $\del$ as specified by the uncertainty set
$$\Delta = \Big\{ \del \; : \; \E_n\left[|\delta_k|^2 \right] \: \leq \: \frac{1}{n}\E_n\left[| \widehat{\rz}\phi_k(\sps) |^2 \right] , \: \forall k  \Big\}.$$ Using \eqref{eq:min-max-opt}, we ensure that the estimated effect is robust against the worst deviations in the set. 

To provide further insight to \eqref{eq:min-max-opt}, we may reformulate the minimax problem as
\begin{equation*}
    \min_{\pvec}  \sqrt{\E_n\Big[ \big| \widehat{\ry} - \widehat{\rz}\basis^\T(\sps)\pvec \big|^2 \Big]} + \sum^{d_\psc}_{k=1} \sqrt{\frac{1}{n}\E_n\left[| \widehat{\rz}\phi_k(\sps) |^2\right]} 
    |\psc_i| ,
\end{equation*}
using Theorem~1 in \citet{xu2009robust}. Note that the first term is the square root of the LS criterion and the second term results in a data-adaptive regularization that mitigates overfitting and biases the estimator against spurious effects. Moreover, the problem can be solved in runtime $\mathcal{O}(n d^2_\psc)$ using the algorithm by \citet{zachariah2015online}. This property enables the efficient computation of bootstrap confidence intervals for $\effectest(\sps)$. We use the pivotal bootstrap method \cite{wasserman2013all} to obtain CIs below.  

\subsection{Estimating residuals}
The conditional means in \eqref{residuals} are MSE-optimal predictors of the exposure and the outcome. In lieu of these, we use predictors $\widehat{\E[\resp|\sps]}$ and $\widehat{\E[\expo|\sps]}$ as approximations. 

Any reasonable prediction method is applicable since \eqref{eq:min-max-opt} mitigates both of the errors $\widetilde{\ry}$ and $\widetilde{\rz}$. To match the form of the spatial model in \eqref{eq:parametricmodel}, we consider here predictors that are linear in the parameters so that
\begin{subequations} \label{estimating residuals}
    \begin{align}
        \widehat{\ry}(\rvec)&=\resp-\basisr^\T(\sps)\rvec,\\ \widehat{\rz}(\rvec)&=\expo-\basisr^\T(\sps)\rvec,
    \end{align}
\end{subequations}
where $\rvec$ denote the parameters and $\basisr(\sps)~=~\col\{1,\basis(\sps)\}$ for simplicity. To learn the parameters, we use here the method in \citet{zachariah2017online} for which learned parameters are given as
\small
\begin{equation*}
    \begin{split}
        {\rvec}_\ry&=\argmin_{\rvec} \sqrt{\E_n\left[\big|\widehat{\ry}(\rvec)\big|^2\right]}+\sum_{k=2}^{d_\psc+1}\sqrt{\frac{1}{n}\E_n\left[|\psi_{k}(\sps)|^2\right]}|\rsc_k|,\\
        {\rvec}_\rz&=\argmin_{\rvec} \sqrt{\E_n\left[\big|\widehat{\rz}(\rvec)\big|^2\right]}+\sum_{k=2}^{d_\psc+1}\sqrt{\frac{1}{n}\E_n\left[|\psi_{k}(\sps)|^2\right]}|\rsc_k|,
    \end{split}
\end{equation*}

\normalsize
where weighted penalty term yields tuning-free regularization that mitigates overfitting in a data-adaptive manner. In addition, $\rvec_\ry$ and $\rvec_\rz$ can be obtained in a computationally efficient manner. Plugging in $\rvec_\ry$ and $\rvec_\rz$ in \eqref{estimating residuals} yields $\widehat{\ry}$ and $\widehat{\rz}$, respectively. 

Combining the predictors with \eqref{eq:min-max-opt} leads to a \emph{r}obust \emph{o}rthogonalized method for \emph{s}patially varying \emph{c}ausal \emph{e}ffect estimation (\methodname).

\subsection{Spatial basis function}
The form of the spatial basis vector $\basis(\sps)$ in \eqref{eq:parametricmodel} depends on whether the space $\sps$ is continuous or discrete. 

\subsection*{Continuous space}\label{continuous basis}
In the continuous case, we consider $\mathcal{S} \subset \mathbb{R}^d$. For clarity of explanation we restrict our description here to the case when $d=2$, but modifications to one or three dimensional space are straightforward. In two dimensions, let $\sps~=~\{\spssc_1,\spssc_2\}$ and
\begin{equation} \label{bspline 2D}
    \basis(\sps)=\basis_1(\spssc_1)\otimes\basis_2(\spssc_2),
\end{equation}
where
\begin{equation*}
\begin{split}
    \basis_1(\spssc_1)&=\col\{\basissc_{i,1}(\spssc_1),\ldots,\basissc_{i,N_{\spssc}}(\spssc_1)\},\\
    \basis_2(\spssc_2)&=\col\{\basissc_{i,1}(\spssc_2),\ldots,\basissc_{i,N_{\spssc}}(\spssc_2)\}.
\end{split}
\end{equation*}
Each $\basis_i(\spssc_i)$ is composed of $N_{\spssc}$ components. Based on their attractive approximation and computational properties, we use the cubic b-spline functions in \eqref{eq:bspline}, where $c$ determines the location of each component and $L$ determines the function support \cite{wasserman2013all}. 
\begin{table*}[!t] 
\centering
\begin{minipage}{0.7\textwidth}
\begin{equation}
  \basissc(s) =
  \begin{cases}
   \frac{1}{6}\bfunc(s)^3 & \text{$\frac{(c-2)L}{4}\leq s <\frac{(c-1)L}{4}$}\\
   \frac{-1}{2}\bfunc(s)^3+2\bfunc(s)^2-2\bfunc(s)+\frac{2}{3} & \text{$\frac{(c-1)L}{4}\leq s <\frac{Lc}{4}$}\\
   \frac{1}{2}\bfunc(s)^3-4\bfunc(s)^2+10\bfunc(s)-\frac{22}{3} & \text{$~~~~~~~\frac{Lc}{4}\leq s <\frac{(c+1)L}{4}$}\\
   \frac{-1}{6}\bfunc(s)^3+2\bfunc(s)^2-8\bfunc(s)+\frac{32}{3} & \text{$\frac{(c+1)L}{4}\leq s \leq\frac{(c+2)L}{4}$}\\
   0 & \text{otherwise}
  \end{cases}
\quad \text{where} \quad \bfunc(s) = \frac{4s}{L} - c+2 
\label{eq:bspline}
\end{equation}
\medskip
\hrule
\end{minipage}
\end{table*}

Thus the effect $\effect(\sps)$ is approximated as a linear combination of b-splines placed at different points across space via the model class \eqref{eq:parametricmodel}. The maximum value of $N_s$ is practically limited by the number of data points $n$ and the available computational resources. For $2$-dimensional space with $N_\spssc$ components the dimension of parameter $\mbs{\theta}$ in \eqref{eq:parametricmodel} is $d_\psc~=~N_{\spssc}^2$. 

The approximation accuracy of the model class introduced above is readily refined by incorporating b-splines with multiple supports. For instance,
\begin{equation} \label{basis multiple support}
    \basis(\sps)=
    \begin{bmatrix}
    \basis_{L_1}(\sps)\\
    \basis_{L_2}(\sps)
    \end{bmatrix}
\end{equation}
accommodates two different support sizes $L_1$ and $L_2$. This enables inferring the effects at different spatial resolutions.

\subsection*{Discrete space}
In the discrete case, the space consists of $d$ disjoint regions $\mathcal{S}=\{1,\ldots,d\}$ and the spatial basis vector is simply
\begin{equation} \label{discrete basis}
    \basis(\sps)=\col\big\{I(\sps=1),\ldots,I(\sps=d)\big\},
\end{equation}
where $I(\sps=k)$ denotes the indicator function. The above choice of $\basis(\sps)$ yields a model class for with an average effect $\effect(\sps)$ for every discrete region $\sps$.

\section{Experimental Results} \label{experimental results}

In this section, we demonstrate the proposed \methodname{} method using simulated data for both continuous and discrete space. We subsequently apply the method to real data.

\subsection{Continuous two-dimensional space}
\begin{figure*}
\centering
\begin{subfigure}{0.24\linewidth}
\includegraphics[width=0.9\linewidth]{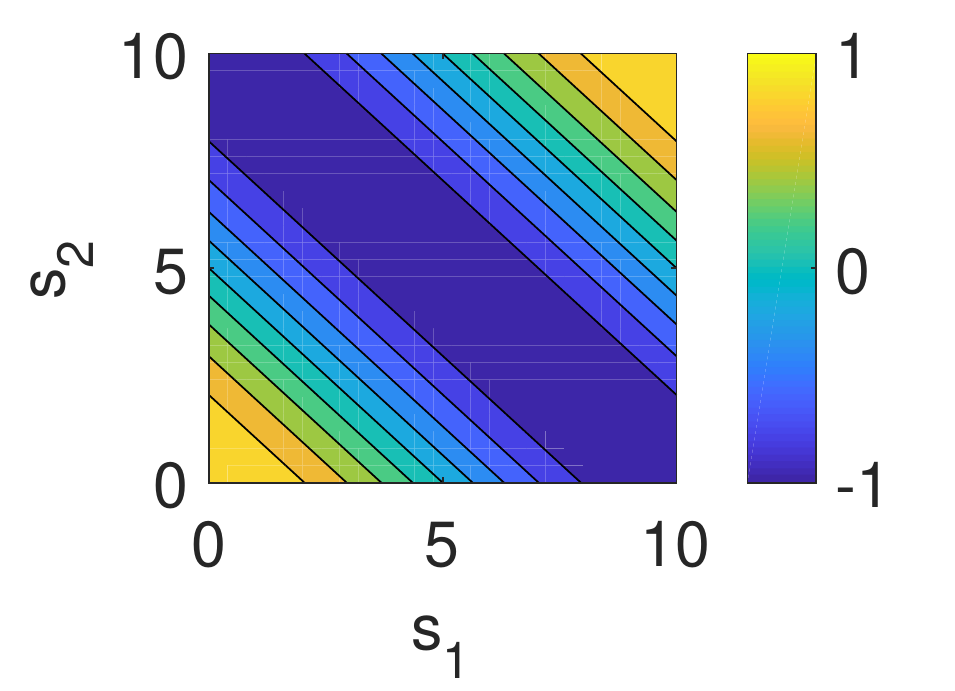}
\caption{}
\label{2D_true}
\end{subfigure}
\begin{subfigure}{0.24\linewidth}
\includegraphics[width=0.9\linewidth]{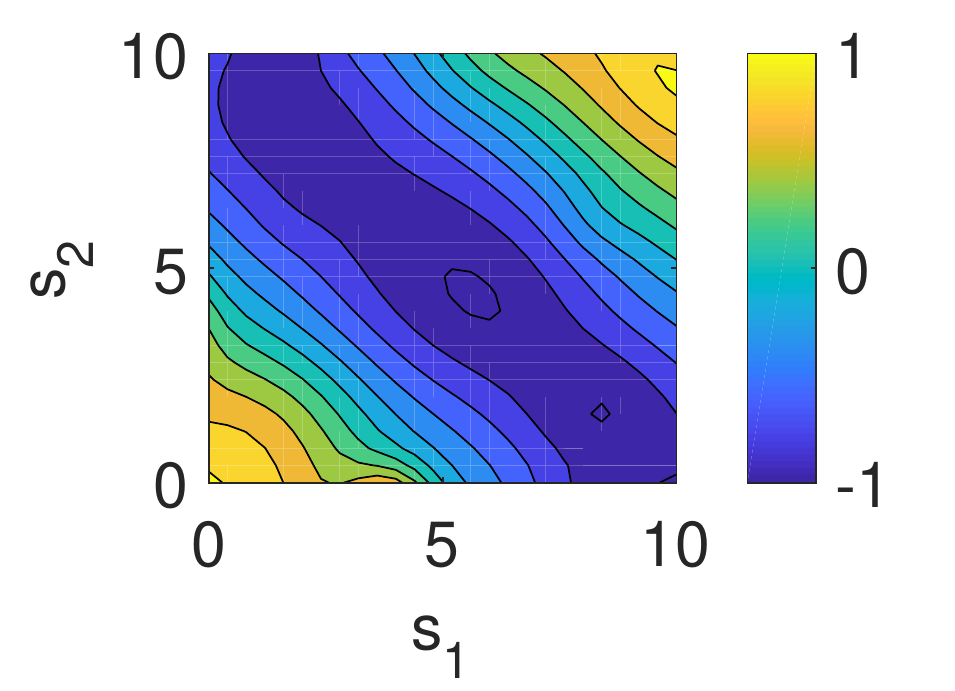}
\caption{}
\label{2D_eff_aprx}
\end{subfigure}
\begin{subfigure}{0.24\linewidth}
\includegraphics[width=0.9\linewidth]{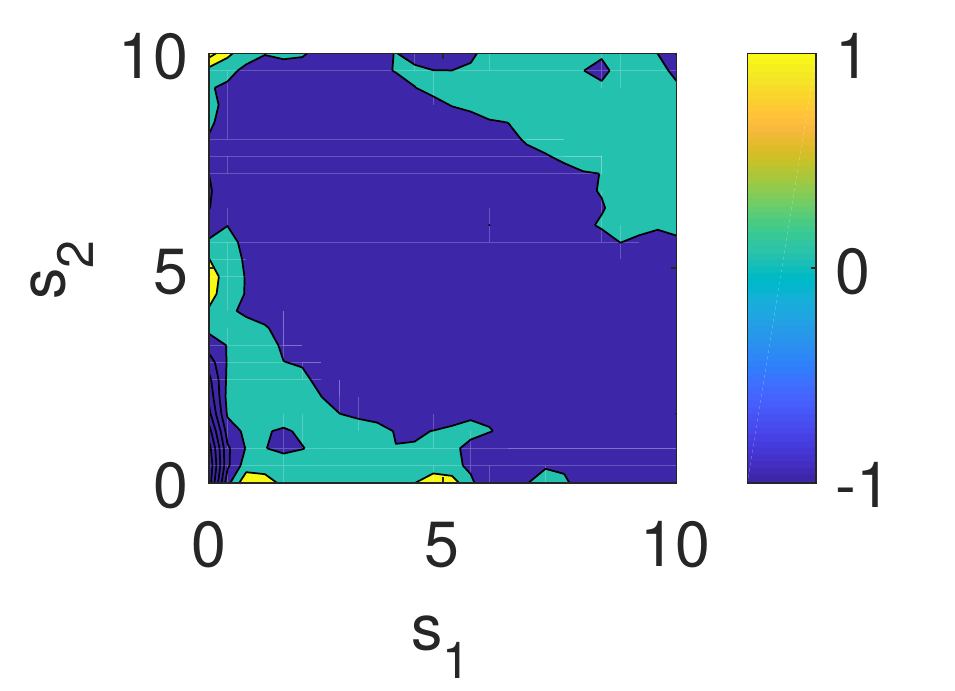}
\caption{}
\label{2D_eff_LS}
\end{subfigure}
\begin{subfigure}{0.24\linewidth}
\includegraphics[width=0.9\linewidth]{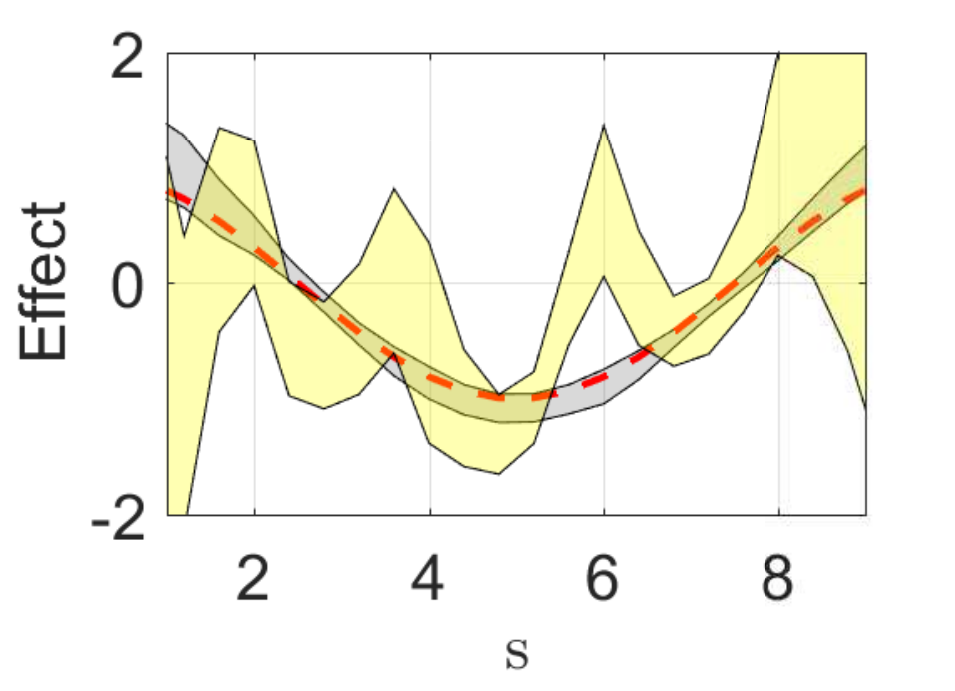}
\caption{}
\label{2D_diag}
\end{subfigure}
\caption{(a) Exposure effect $\effect(\sps)$ across two-dimensional space $\mathcal{S}$ and estimated effects $\effectest(\sps)$ from $\dataset_n$ using (b) \methodname{} and (c) direct approach. (d) $\effect(\sps)$ along left diagonal (dashed line) along with 95\%-CI for \methodname{} (grey shaded) and direct approach (yellow shaded).  }
\label{2D_cont}
\end{figure*}

First, we illustrate \methodname{} in continuous two-dimensional space $\mathcal{S} = [0,10]^2$. The outcome is generated according to 
\begin{equation*}
    \resp=\effect(\sps)\expo+\beta(\sps)+\varepsilon,
\end{equation*}
where the effect of the exposure varies according to
\begin{equation*}
    \begin{split}
    \effect(\sps)&=\cos{\left(\frac{2\pi\spssc_1}{20}+\frac{2\pi\spssc_2}{20}\right)},
    \end{split}
\end{equation*}
as shown in Figure~\ref{2D_true}. Note that $\effect(\sps)$ does not belong to the model class \eqref{eq:parametricmodel}, but can nevertheless be well approximated using the b-splines basis. The nuisance function is $\beta(\sps)=\basis^\T_0(\sps) \mbs{\eta}$, where $\basis_0(\sps)$ is a b-spline basis with $N_s = 10$ and support $L = 0.2\times10$ ($10$ being the range of $\sps$ in each dimension), see Sec.~\ref{continuous basis}. The unknown coefficients were drawn as $\mbs{\eta} \sim \mathcal{N}(\mathbf{0},\mathbf{I})$.  The exposure and error are distributed as $\expo \sim \mathcal{N}( 0.5 \times \beta(\sps), 1 )$ and $\varepsilon \sim \mathcal{N}(0,0.2^2)$, respectively. We generate a dataset $\mathcal{D}_n~=\{(\resp_i,\expo_i,\sps_i)\}_{i=1}^{n}$, where $n=676$. The samples of $\sps$ were drawn uniformly across $\mathcal{S}$.

For estimation, we use a basis vector $\basis(\sps)$ with $N_s = 10$. To capture multiple resolutions, we use three levels of supports $L_1~=~0.2\times10,~L_2~=~0.4\times10$ and $L_3~=~0.85\times10$, cf. \eqref{basis multiple support}. For the sake of comparison, we consider a direct approach to estimate $\effect(\sps)$ using \eqref{eq3}, which requires the joint estimation of the nuisance function $\beta(\sps)$. For the latter, we use the well-specified model $\basis^\T(\sps)\rvec$, which contains $\basis_0(\sps)$ since $L_1 = L$. The effect and nuisance functions are jointly estimated using the LS method.


Figure~\ref{2D_eff_aprx} and \ref{2D_eff_LS} show the estimate $\widehat{\effect}(\sps)$ obtained using \methodname{} and the direct approach, respectively. It can be clearly seen that the latter results in very poor effect estimates, due to the high-dimensionality of $\beta(\sps)$. In contrast to this, the \methodname{} method approximates the true effect well. Moreover, we also evaluate $95\%$ confidence interval (CI) using the pivotal bootstrap method \cite{wasserman2013all} and $3000$ bootstrap iterations for both methods. Figure~\ref{2D_diag} shows the effect $\effect(\sps)$ along the left diagonal with the $95\%$ CIs of the two methods. The direct approach infers insignificant effects, both where the true effect is actually positive and as well where the true effect is negative. The inference is also inconsistent, reporting a significant effect with an opposite sign to the true effect at around $s = 6$. The CI of \methodname{}, on the other hand, is consistent with the true effect. 


\subsection{Discrete spatial regions}
Next, we illustrate  \methodname{} in the simple case of discrete space $\mathcal{S}=\{1, \dots, d\}$ and compare it with a naive LS approach. The outcome is generated according to 
\begin{equation*}
    \resp=\effect(\sps)\expo+\beta(\sps)+\varepsilon,
\end{equation*}
where the effect of the exposure varies across $d=5$ regions as
\begin{equation*}
    \effect(\sps)=\cos{\left(\frac{2\pi}{2d}\sps\right)}
\end{equation*}
and the nuisance function is $\beta(\sps) = 2-\sps$. The exposure and error are distributed as $\expo \sim \mathcal{N}(\cos{\left(\frac{2\pi}{2d}\sps\right)},1)$ and $\varepsilon \sim \mathcal{N}(0, 0.2^2)$, respectively. We generate a dataset $\mathcal{D}_n$, where $n=500$ and the samples are drawn uniformly across space.

\begin{figure}[!t]
    \centering
    \begin{subfigure}{0.75\linewidth}
    \includegraphics[width=0.9\linewidth]{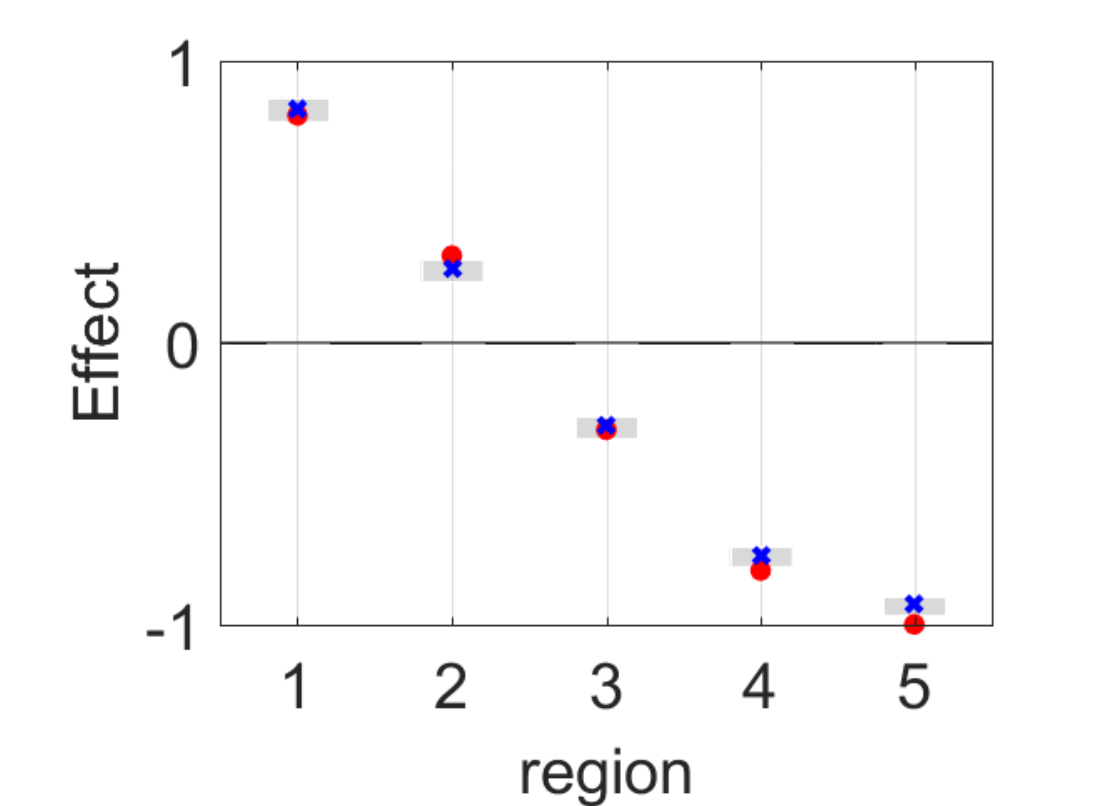}
    \caption{}
    \label{dis_cosine_gam_our_model}
    \end{subfigure}
    \begin{subfigure}{0.75\linewidth}
    \includegraphics[width=0.9\linewidth]{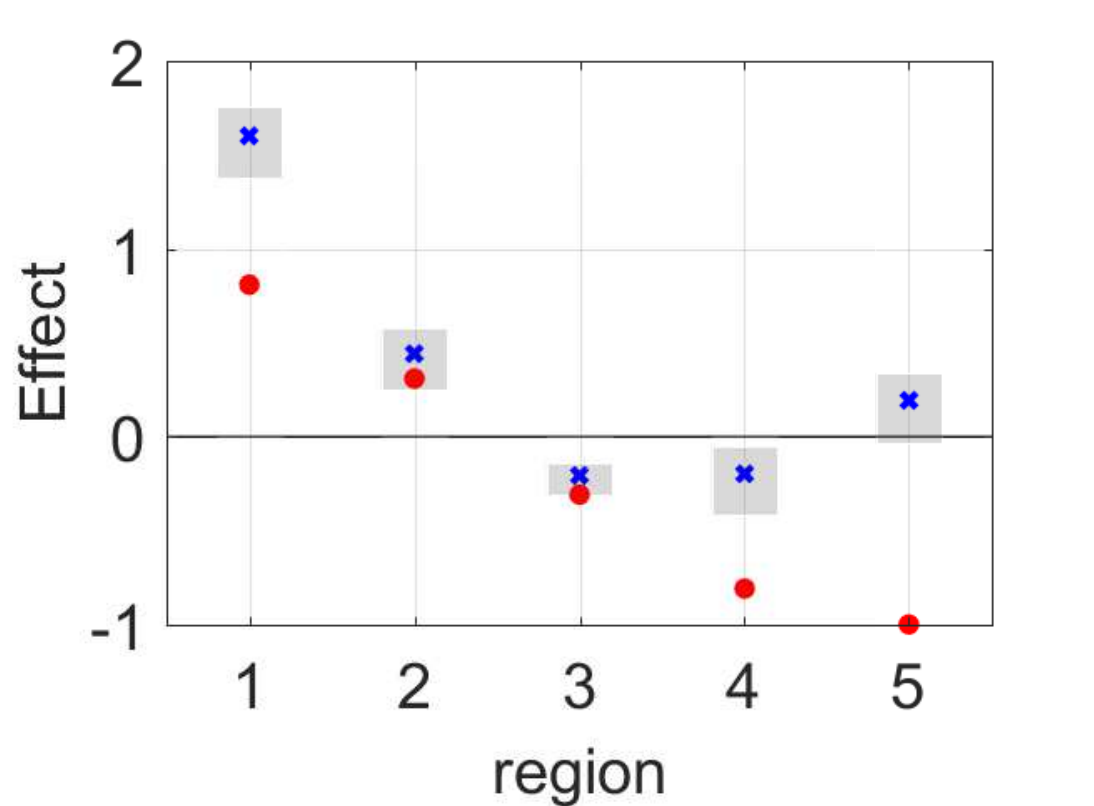}
    \caption{}
    \label{dis_cosine_gam_naive_LS}
    \end{subfigure}
    \caption{Effect $\effect(\sps)$ (dots), estimated effect $\widehat{\effect}(\sps)$ (crosses) and the 95\%-CI (shaded) across five spatial regions using (a) \methodname{} (b) standard regression approach.} 
    \label{hetr_disc}
\end{figure}

The vector $\basis(\sps)$ is given by \eqref{discrete basis}. For the sake of comparison, we consider a standard LS regression approach, in which the predictive effect of the exposure in each region is estimated separately. Figures \ref{dis_cosine_gam_our_model} and \ref{dis_cosine_gam_naive_LS} illustrate the difference between \methodname{}, which takes the causal structure into account, and a standard regression approach. The latter leads to nonnegligible biases and misleading inferences, see for instance the inferred effect in region 5.

\subsection{Robustness to errors-in-variables}
Finally, we illustrate the robustness of the method against the unobserved deviations $\del(\sps)$. We consider one-dimensional continuous space $\mathcal{S} = [0,10]$ with an effect parameterized as $\effect(\sps) = \basis^\T(\sps)\pvec_0$, where $\pvec_0$ has dimension $d_{\psc} = 10$. We generate $n=41$ samples and, for clarity, let $\widetilde{\varepsilon}$ in \eqref{eq:residualregression} be zero, so that the only estimation error arises via $\widetilde{\rz}$. The residual $\rz~=~\widehat{\rz}~+~\widetilde{\rz}$ in \eqref{residuals} is generated with independent variables
$$\widehat{\rz}, \: \widetilde{\rz} \: \sim \mathcal{N}(0,1),$$
which yields large errors.

\begin{figure}[!t]
    \centering
    \begin{subfigure}{0.75\linewidth}
    \includegraphics[width=0.9\linewidth]{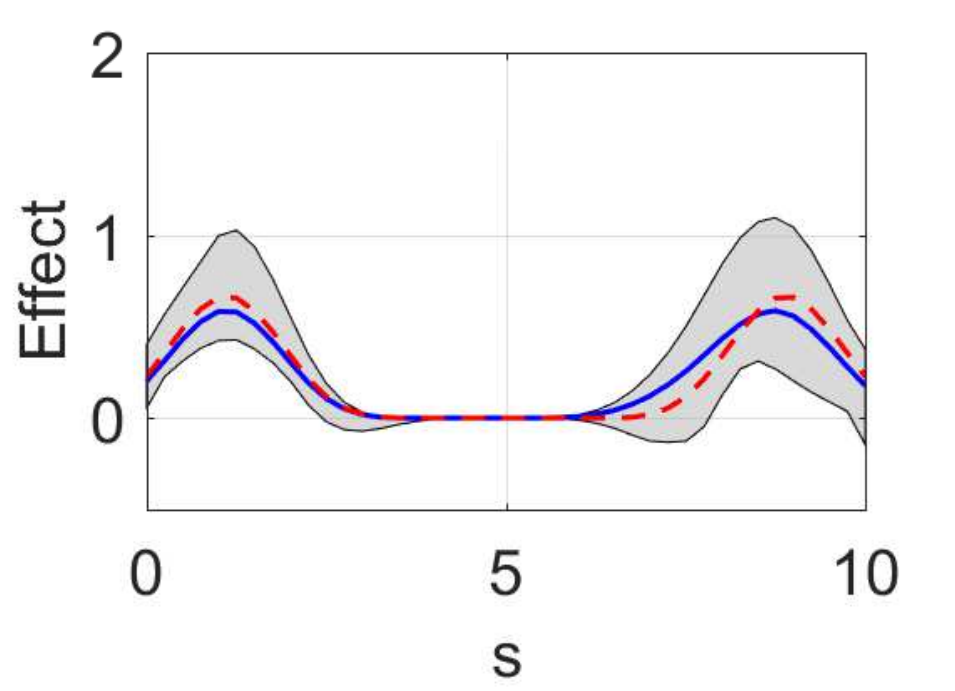}
    \caption{}
    \label{robust_ci}
    \end{subfigure}
    \begin{subfigure}{0.75\linewidth}
    \includegraphics[width=0.9\linewidth]{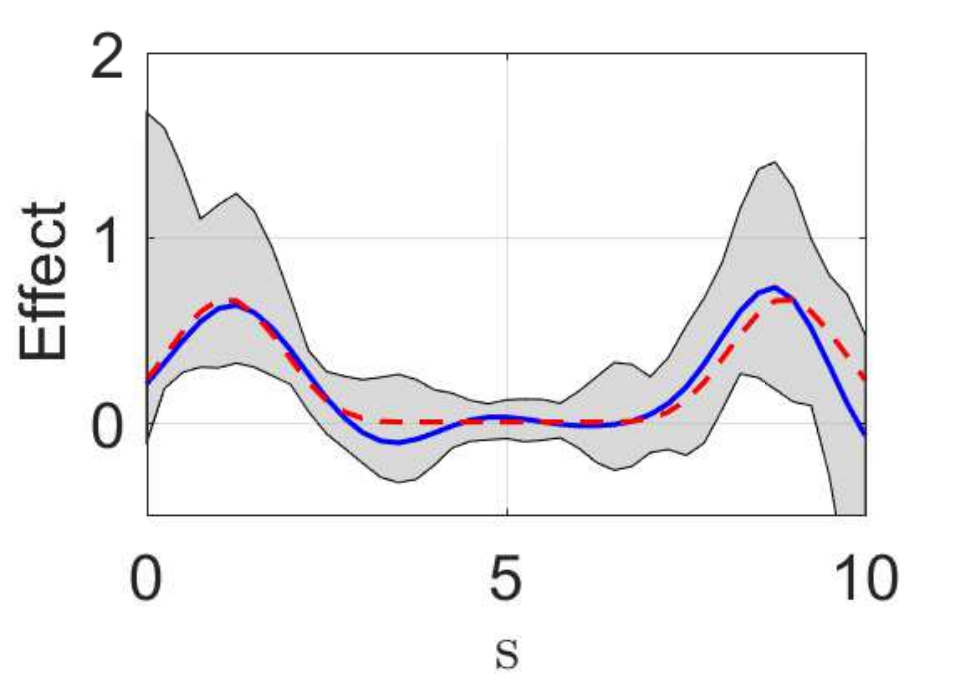}
    \caption{}
    \label{robust_ci_ls}
    \end{subfigure}
    \begin{subfigure}{0.75\linewidth}
    \includegraphics[width=0.9\linewidth]{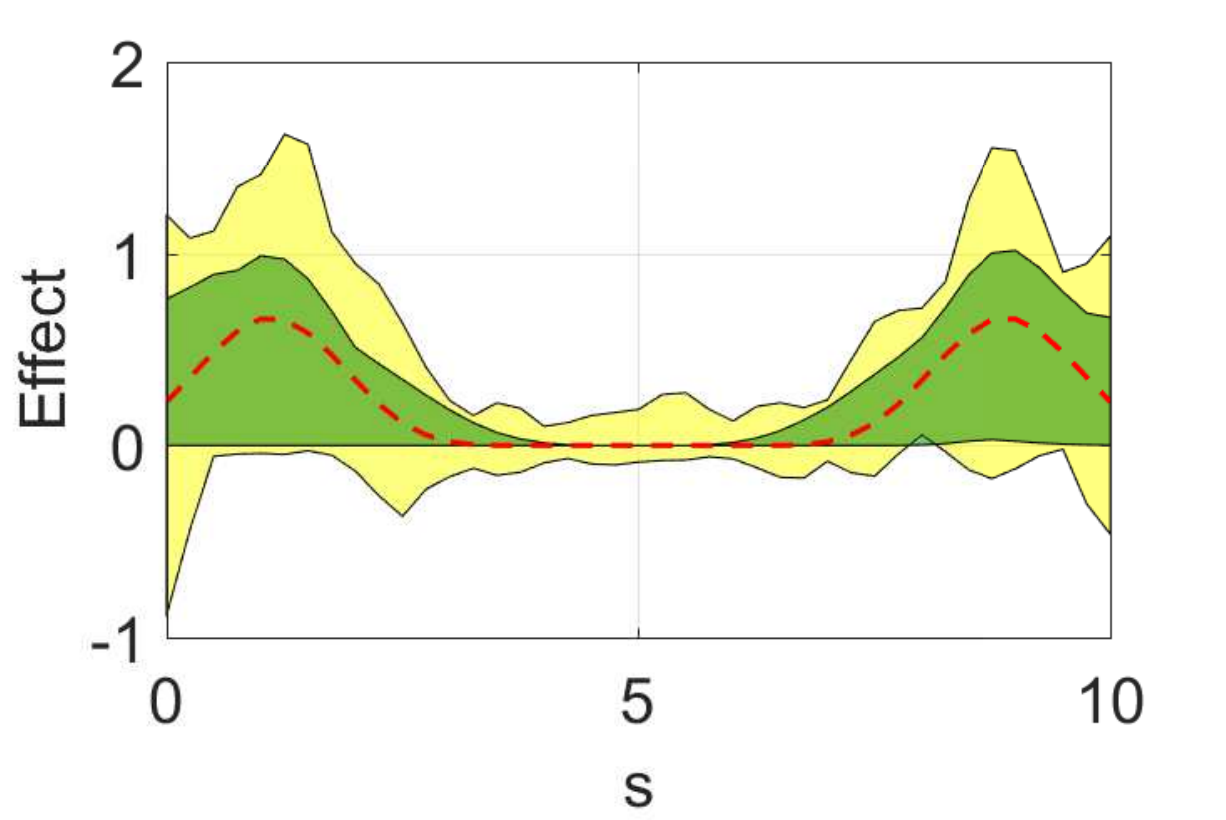}
    \caption{}
    \label{robust_msims}
    \end{subfigure}
   
    \caption{Illustrating robustness to errors-in-variables. True effect $\effect(\sps)$ (dashed line), estimate $\effectest(\sps)$ (solid line) and 95\%-CI (grey shaded region) using (a) \methodname{} and (b) LS-based method for single dataset. (c) Dispersion of estimates using \methodname{} (green shaded) and LS (yellow shaded) in terms of $5^\text{th}$ and $95^\text{th}$ quantiles $[Q_{5},Q_{95}]$, based on $100$ Monte Carlo simulations. Note where the estimates are positive, \methodname{} concentrates the estimates on the positive side for all simulations in contrast to the LS method.}
    \label{robust}
\end{figure}

We compare \methodname{} with an LS estimator applied to \eqref{eq:residualregression}, which assumes $\del \equiv \mathbf{0}$. An example realization of the estimates $\effectest(\sps)$ is shown in Figures \ref{robust_ci} and \ref{robust_ci_ls}. We see that \methodname{} biases the estimates towards zero and reduces the risk of reporting spurious effects. Correspondingly, this method reports larger regions in which effects are significant (via 95\%-CIs) as compared to the LS method. We repeat the experiment using Monte Carlo simulations and confirm in Figure \ref{robust_msims} that the proposed method is less prone to report spurious effects under errors-in-variables than an LS-based approach.


\subsection{German income data} 

We proceed to illustrate the method with real datasets. In these examples, the space is partitioned into discrete regions. First, we consider the average household income in Euros at county level in Germany for the year $2012$ \cite{Germandata_source}.

\begin{figure*}[!t]
    \centering
    \begin{subfigure}{0.2\linewidth}
    \includegraphics[width=0.8\linewidth]{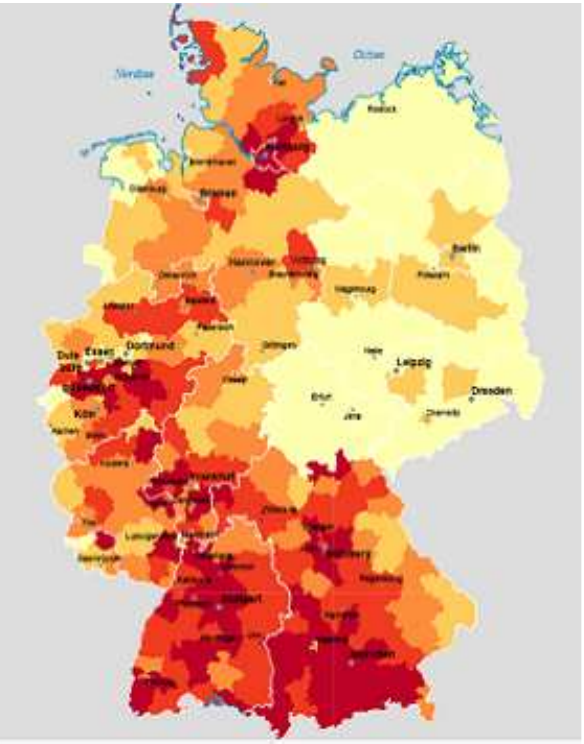}
    \caption{}
    \label{income}
    \end{subfigure}
    \centering
    \begin{subfigure}{0.2\linewidth}
    \includegraphics[width=0.8\linewidth]{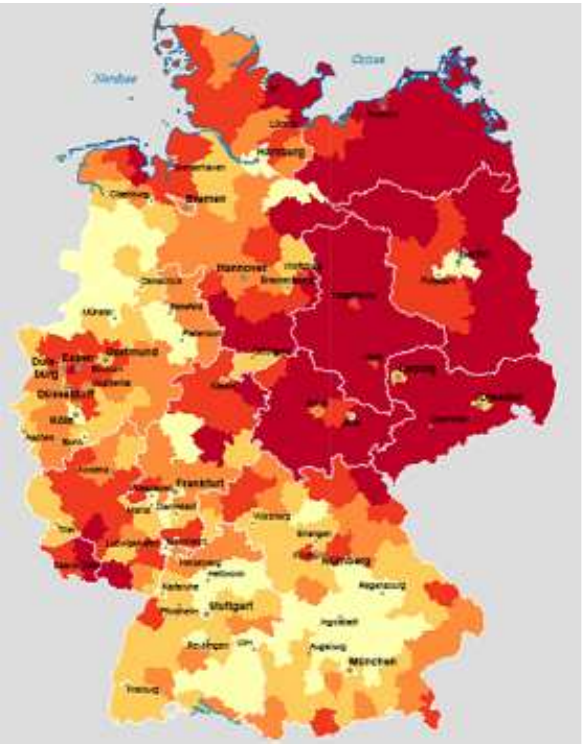}
    \caption{}
    \label{age}
    \end{subfigure}
    \centering
    \begin{subfigure}{0.2\linewidth}
    \includegraphics[width=0.8\linewidth]{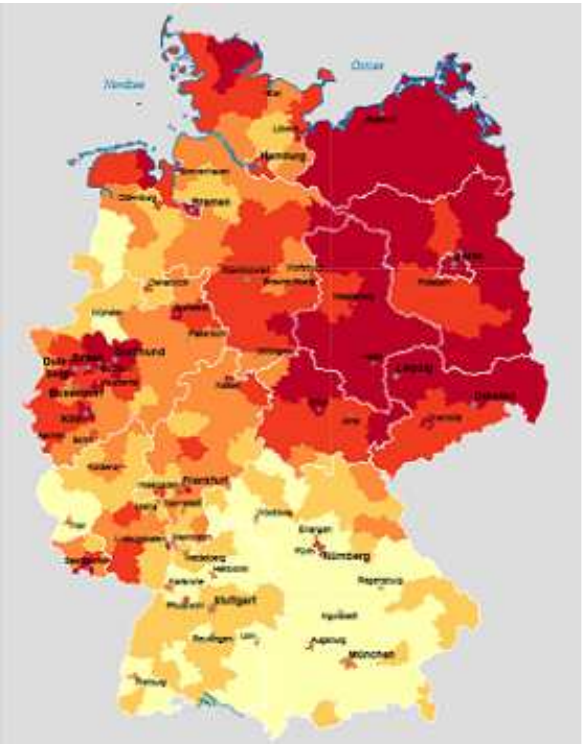}
    \caption{}
    \label{unemploy}
    \end{subfigure}
    \caption{German data (a) Income [Euro]. (b) Age [years]. (c) Unemployment rate [$\%$]. The dark color denotes high values while the light color denotes small values.}
    \label{germ_data}
\end{figure*}

Suppose we are interested in studying the effect of age $\expo$ on income $\resp$ across Germany's federal states. The observed data is $\mathcal{D}_n~=~\{(\resp_i,\expo_i,\sps_i)\}_{i=1}^{n=366}$ which spans  $d~=~11$ states. The data is standardized to have zero mean and unit standard deviation. For the sake of illustration, suppose that different regions have different age demographics but also different economic structures with varying levels of unemployment. Then the unemployment level, denoted $\cnf$, acts as a spatial confounder on the relation between age and income \cite{thaden2018structural}, see Figure~\ref{germ_data}.

\begin{figure}[!t]
    \centering
    \begin{subfigure}{0.8\linewidth}
    \includegraphics[width=0.85\linewidth]{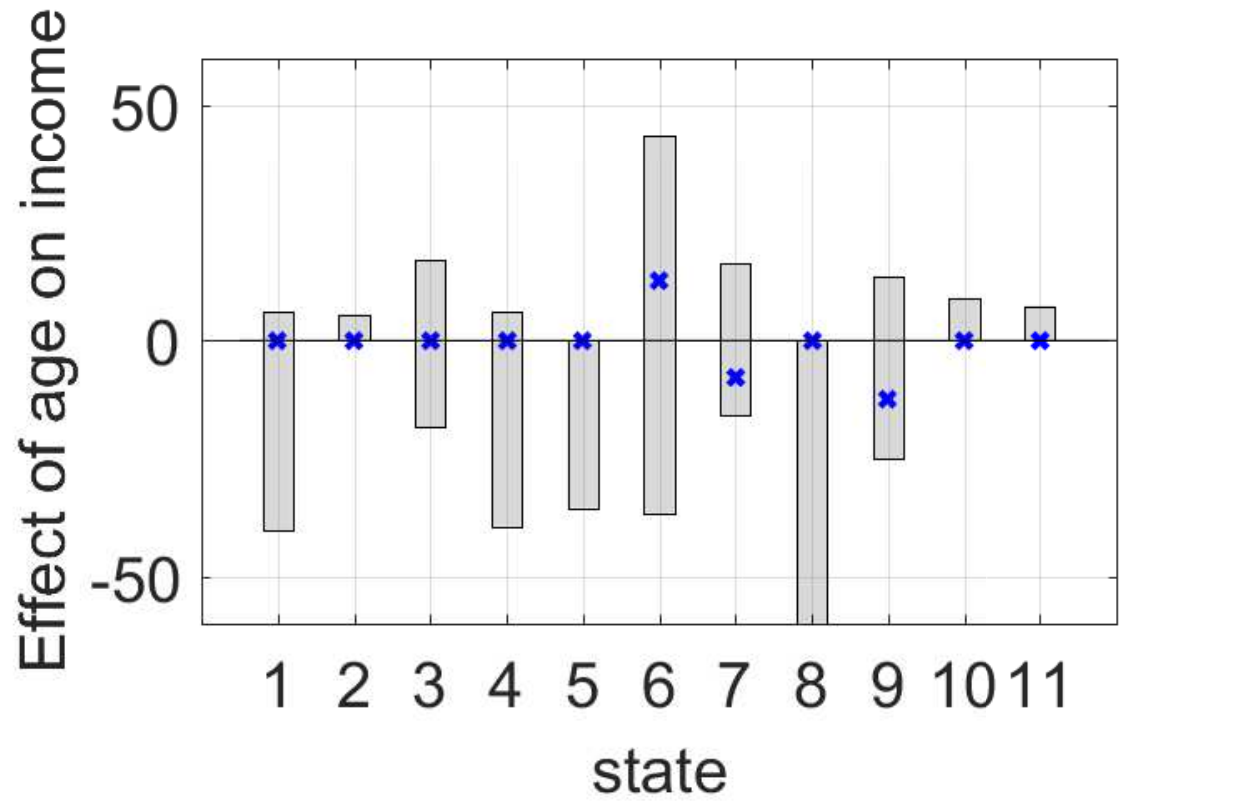}
    \caption{}
    \label{germ_age_empl_hid}
    \end{subfigure}
    \centering
    \begin{subfigure}{0.8\linewidth}
    \includegraphics[width=0.9\linewidth]{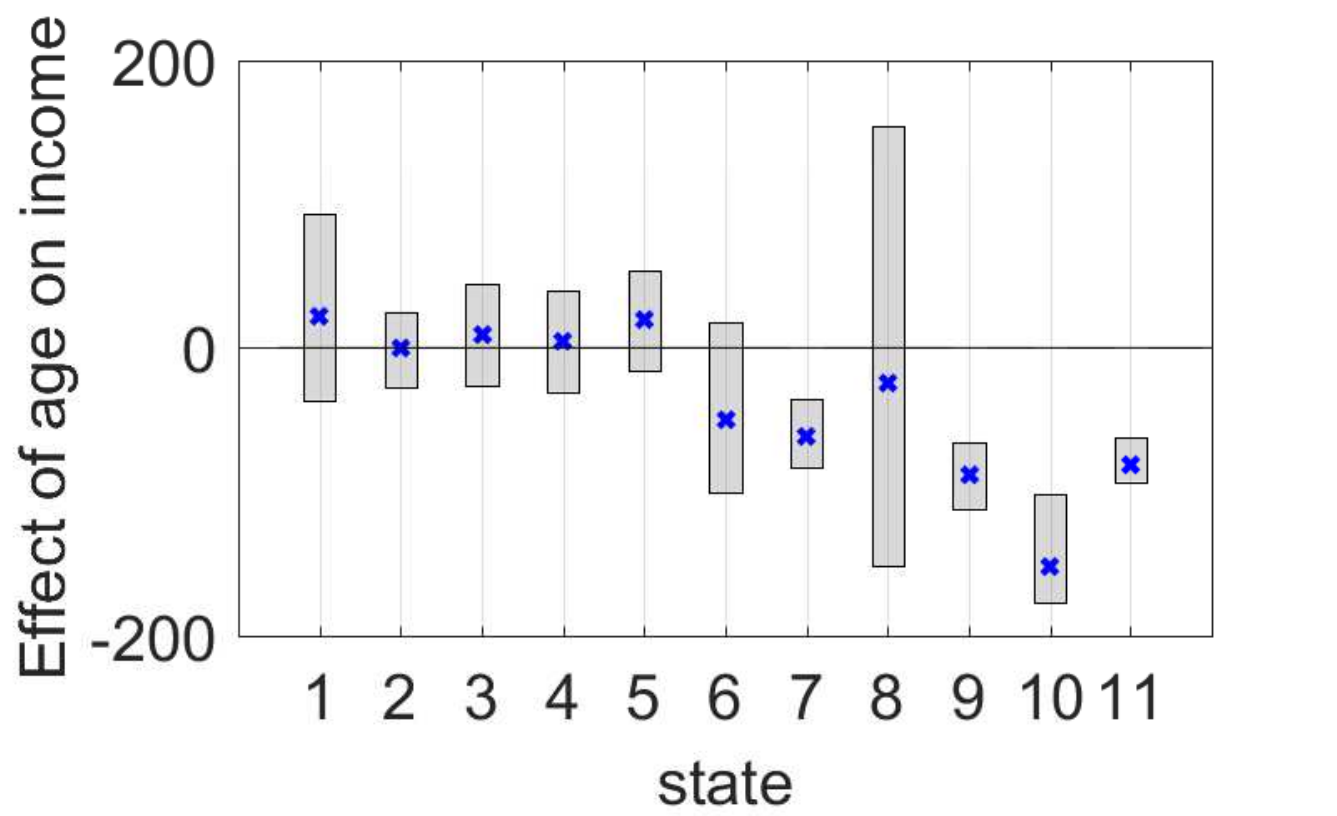}
    \caption{}
    \label{germ_naive}
    \end{subfigure}
    \centering
    \begin{subfigure}{0.8\linewidth}
    \includegraphics[width=0.9\linewidth]{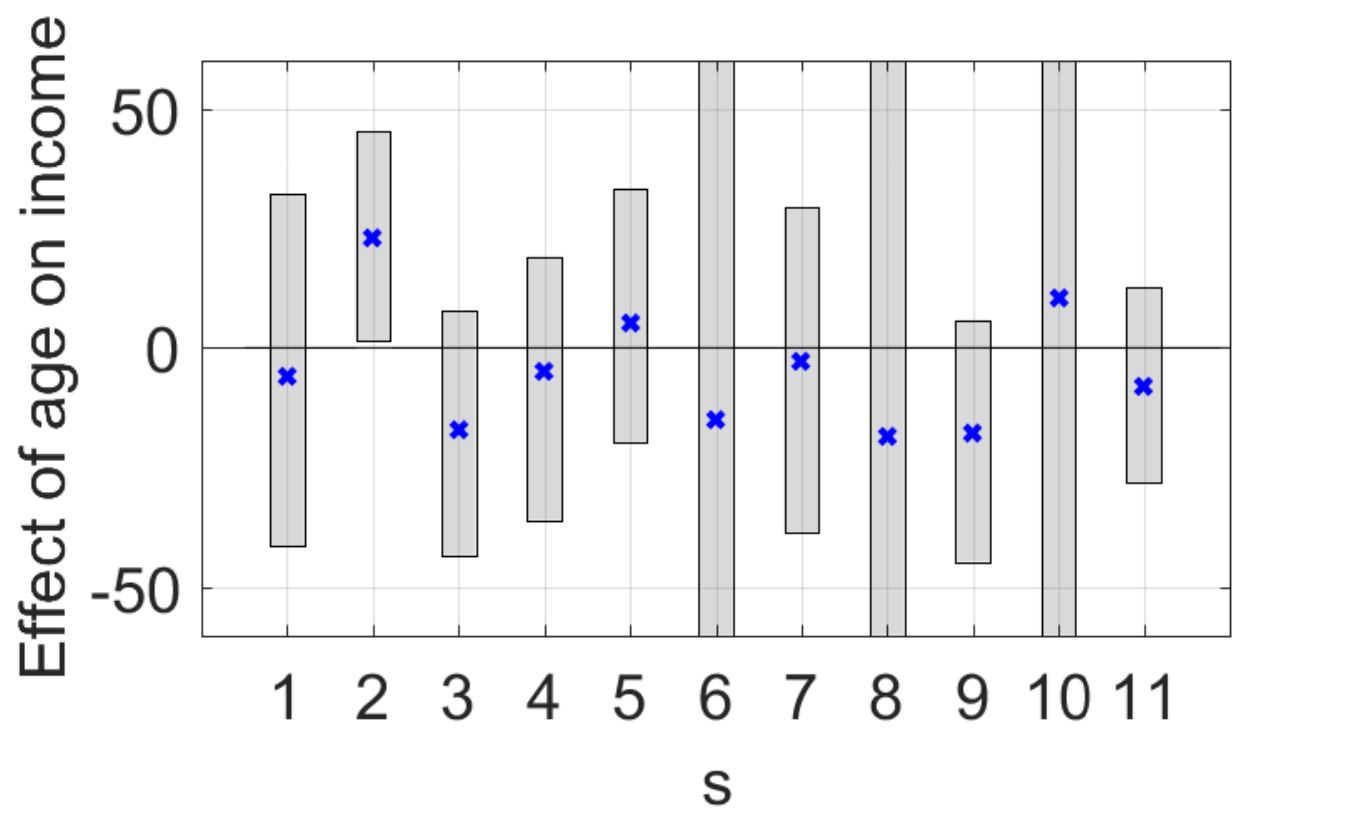}
    \caption{}
    \label{germ_age_empl_incl}
    \end{subfigure}
    \caption{Inferred effect of age on median income $\widehat{\effect}(\sps)$ (blue crosses) along with 95\%-CI (shaded) for $11$ different federal states in Germany. (a) \methodname{}, (b) standard regression approach, (c) regression after controlling for unemployment.} 
    \label{germ}
\end{figure}

Assuming that $\cnf$ is unknown, we use \methodname{} to infer the average change in income per year of age, i.e., $\tau(\sps)$. The result is shown in Figure~\ref{germ_age_empl_hid}, where we do not infer any significant causal effects across states. By contrast, when using a standard regression approach to infer the predictive effect in each state, we find significant negative effects of age in four states, see Figure \ref{germ_naive}. For comparison with \methodname{}, we control for county-level unemployment rates in each state by partialing out $\cnf$ from both $\expo$ and $\resp$, and then estimating the effect from the residuals. The results in Figure~\ref{germ_age_empl_incl} corroborate those reported in Figure~\ref{germ_age_empl_hid} in that the effects remain insignificant when the unobserved variables are taken into account.

\subsection{US Crime data}

Next, we are interested in studying the effect of poverty on crime, which has been a longstanding research topic in criminology. Based on a review of $273$ different studies, \citet{ellis2001crime} concluded that there is an inverse association between an individual's socioeconomic status and criminal behavior. We use number of crimes and number of poor families data on county level from US census of year $2000$ \cite{USAdata_source} to test if our method is able infer such a causal effect. The outcome $\resp$ considered is the total number of crimes due to theft, larcency and burglary. The exposure $\expo$ considered is the number of families that are classified as poor.  The observed data is $\mathcal{D}_n~=~\{(\resp_i,\expo_i,\sps_i)\}_{i=1}^{n=2786}$ which spans over $d~=~50$ states.

\begin{figure}[!t]
    \centering
    \begin{subfigure}{0.8\linewidth}
    \includegraphics[width=0.9\linewidth]{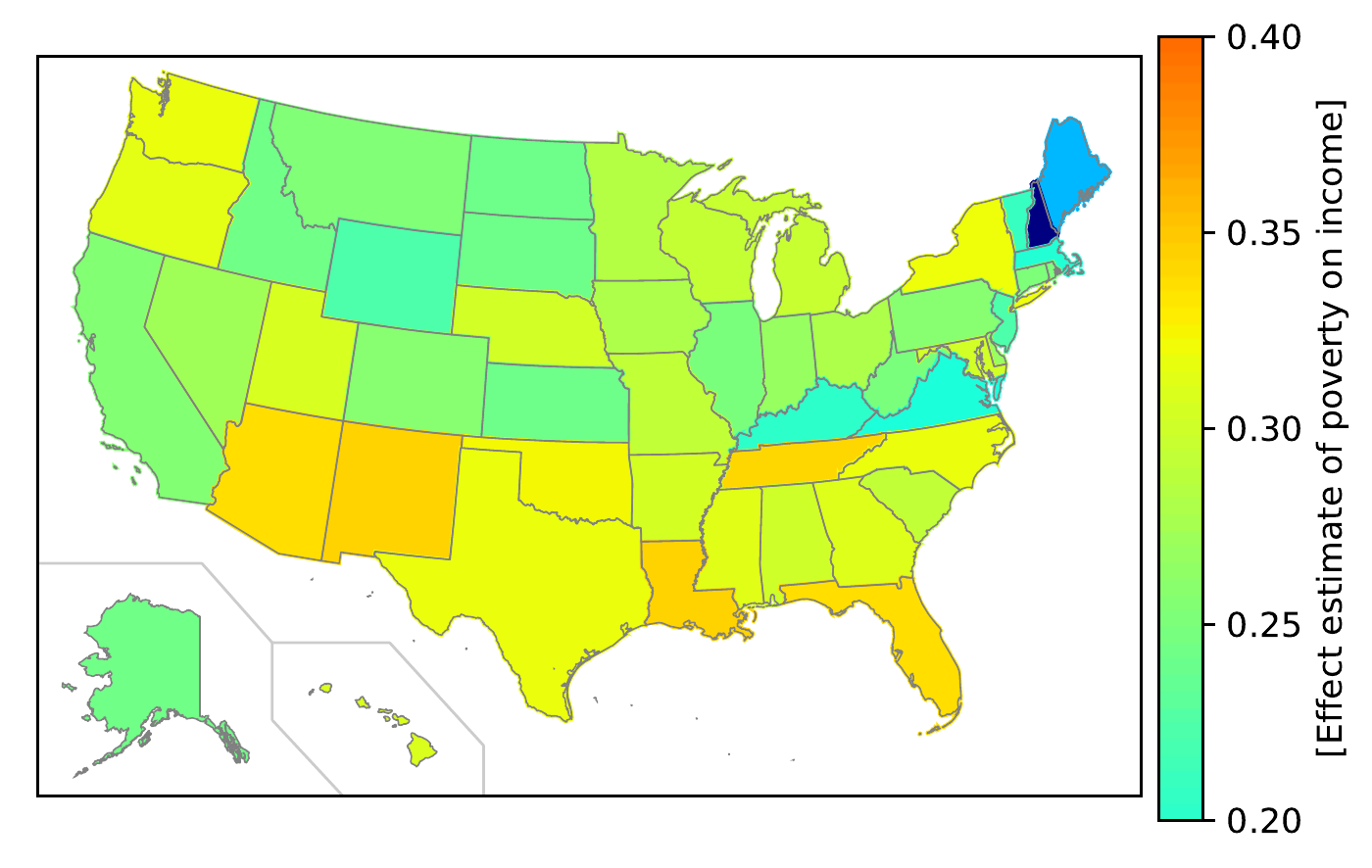}
    \caption{}
    \label{us_pvy}
    \end{subfigure}
    \begin{subfigure}{0.8\linewidth}
    \includegraphics[width=0.9\linewidth]{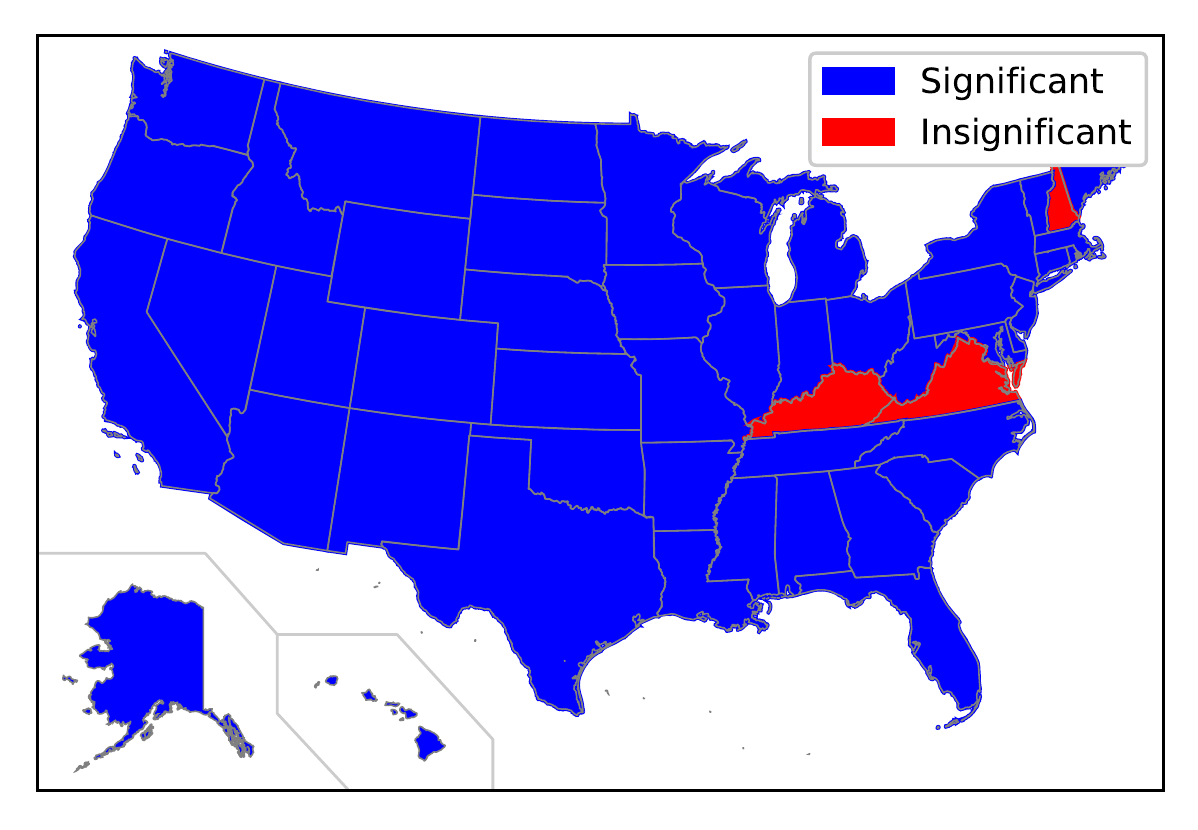}
    \caption{}
   \label{us_pvy_sig}
    \end{subfigure}
    
    \caption{USA data: (a) The estimated effect $\effectest(\sps)$ of number of poor families on average number of crimes due to theft, larcency and burglary. (b) Significance of estimated effect $\effectest(\sps)$. }
    \label{us_pvy}
\end{figure}

Figure \ref{us_pvy} shows the estimated effect of number of poor families on crimes due to theft, larcency and burglary $\effectest(\sps)$. It is estimated to be positive for all states, where the estimates are significant in all but three states as shown in Figure \ref{us_pvy_sig}. The results are thus consistent with previous findings. We find that across most states, eliminating poverty among ten families, reduces on average three crimes (of either theft, larceny and burglary).

We also consider the effect of education on income using percentage of people getting high school education and further and median annual household income data at county level from US census of year $2000$ \cite{USAdata_source}. Figure $S1$ in the supplementary material shows the estimated effect $\effectest(\sps)$ of education attainment percentage on annual median household income. Our method gives a positive effect which one would intuitively expect between these two quantities.




\section{Conclusion}
We have proposed a method for estimating heterogeneous effects of an exposure on an outcome variable in presence of possible unmeasured spatially varying confounding variables. The method is applicable to both continuous and discrete space $\sps$. The orthogonalized approach circumvents joint estimation of a possibly high-dimensional nuisance function while the robust estimation approach mitigates the resulting errors-in-variables. We demonstrated the properties of the method on synthetic data and illustrated its potential use in real-world applications. Link to code: \href{https://github.com/Muhammad-Osama/Inferring-Heterogeneous-Causal-Effects-in-Presence-of-Spatial-Confounding.git}{github}.


\clearpage

\bibliography{main}
\bibliographystyle{icml2019}

\section*{Acknowledgements}
The authors would like to thank Fredrik Johansson (MIT) for feedback on an early version of this work. This research was financially supported by the projects \emph{Counterfactual Prediction Methods for Heterogeneous Populations} (contract number: 2018-05040) and \emph{NewLEADS - New Directions in Learning Dynamical Systems} (contract number: 621-2016-06079), funded by the Swedish Research Council and by the Swedish Foundation for Strategic Research (SSF) via the project \emph{ASSEMBLE} (contract number: RIT15-0012).

\end{document}